\documentclass[english,prb, aps, twocolumn]{revtex4}
\usepackage[T1]{fontenc}
\usepackage[latin9]{inputenc}
\usepackage{graphicx}

\makeatletter
\@ifundefined{textcolor}{}
{%
 \definecolor{BLACK}{gray}{0}
 \definecolor{WHITE}{gray}{1}
 \definecolor{RED}{rgb}{1,0,0}
 \definecolor{GREEN}{rgb}{0,1,0}
 \definecolor{BLUE}{rgb}{0,0,1}
 \definecolor{CYAN}{cmyk}{1,0,0,0}
 \definecolor{MAGENTA}{cmyk}{0,1,0,0}
 \definecolor{YELLOW}{cmyk}{0,0,1,0}
 }

\makeatother

\usepackage{babel}
\begin{document}

\title{Universal Response Curve for Nanowire Superconducting Single-Photon
Detectors}

\author{J.J. Renema\textsuperscript{1)}, G. Frucci\textsuperscript{2)},
Z. Zhou\textsuperscript{2)}, F. Mattioli\textsuperscript{3)}, A.
Gaggero\textsuperscript{3)}, R. Leoni\textsuperscript{3)}, M.J.A.
de Dood\textsuperscript{1)}, A. Fiore\textsuperscript{2)}, M.P.
van Exter\textsuperscript{1)}}

\affiliation{1) Leiden Institute of Physics, Leiden University, Niels Bohrweg
2, 2333 CA Leiden, the Netherlands }

\affiliation{2) COBRA Research Institute, Eindhoven University of Technology,
P.O. Box 513, 5600 MB Eindhoven, The Netherlands }

\affiliation{3) Istituto di Fotonica e Nanotecnologie (IFN), CNR, via Cineto Romano
42, 00156 Roma, Italy }
\begin{abstract}
Using detector tomography, we investigate the detection mechanism
in NbN-based superconducting single photon detectors (SSPDs). We demonstrate
that the detection probability uniquely depends on a particular linear
combination of bias current and energy, for a large variation of bias
currents, input energies and detection probabilities, producing a
universal detection curve. We obtain this result by studying multiphoton
excitations in a nanodetector with a sparsity-based tomographic method
that allows factoring out of the optical absorption. We discuss the
implication of our model system for the understanding of meander-type
SSPDs.
\end{abstract}
\maketitle

\section{Introduction}

Nanowire Superconducting Single Photon Detectors (SSPDs) \cite{Goltsman2001}
have high detection efficiency \cite{Marsili}, low dark counts, low
jitter and a broadband absorption spectrum \cite{Verevkin2002}. This
makes them suitable for many applications including quantum optics
\cite{Stevens2010,Zinoni2007,GisinnatureP,RenemaPRA}, quantum key
distribution \cite{Hadfield2006,Collins2007}, optical coherence domain
reflectometry \cite{Mohan2008} and interplanetary communication \cite{Boroson2009}.
These detectors typically consist of a thin nanowire ($\sim$4 nm
x 100 nm) of superconducting material, such as NbN \cite{Goltsman2001},
TaN \cite{Engel2012}, NbTiN \cite{Dorenbos2008}, Nb \cite{Annunziata2009},
or WSi \cite{Marsili}, which is typically fabricated in a meander
shape to cover an active area of 25-1600 $\mu$m$^{2}\;$ \cite{Mattioli2012}.
The absorption of a single photon in the nanowire results in the creation
of a a region with a non-equilibrium concentration of quasiparticles.
When the nanowire is biased close to the critical current, this perturbation
causes a transition from the superconducting to the resistive state,
producing a voltage pulse in the external circuit. 

While progress has been made in understanding the detection process,
many crucial features of the process are still unknown. In this publication,
we investigate the detection process by means of a model system: an
NbN nanodetector \cite{Bitauld2010} (see figure 1). This detector
has a single cross section of wire as its active element, defined
by a bowtie-shaped constriction. We investigate this system with sparsity-based
detector tomography. The tomographic method does not require a model
of the device, which makes it ideally suited as a tool for investigating
the working principle of a detector of which the working mechanism
is not fully understood yet.

It has long been known that at lower bias current, the detector operates
in a regime where multiple photons are necessary to break the superconductivity
\cite{Goltsman2001,Zhang2003}. In a nanodetector, the geometry is
such that many multiphoton processes play a strong role \cite{Bitauld2010,Renema2012}.
This enables us to probe the response of the device to excitations
at different energies simultaneously. The role of detector tomography
is to extract the effects of the various multiphoton excitations. 

In this paper, we investigate the detection process by combining tomography
and a nanodetector. With this combination, we can probe the system
in a way that is independent of the incoupling efficiency of light
into the detector. Moreover, because we tune the energy of the excitation
via the number of photons at the same wavelength, we are insensitive
to wavelength-dependent effects in the setup. This combination allows
us to focus on the fundamentals of the detection process. We demonstrate
that for intrinsic detection probabilities ranging from 0.3 to $10^{-4}$,
the detection probability depends only on a specific linear combination
of bias current and excitation energy. Thus, we obtain a universal
detection curve for our model system of an SSPD: for each bias current
and excitation (photon) energy, the detection probability is given
by a point on this single curve. This universal curve stretches from
the regime where photodetection is almost deterministic (given that
the photon is absorbed into the active area) to the regime where fluctuations
in the wire are thought to play a role in assisting the detection
process.

\section{Experiment}

All experiments in this work were performed on a nanodetector (see
Fig 1). The nanodetector consists of 4 nm thin NbN wire on a GaAs
substrate, shaped into a 150 nm wide bowtie geometry. The device was
fabricated via a combination of DC magnetron sputtering \cite{Gaggero2010},
electron-beam litography, reactive ion etching and evaporation of
the metal contacts \cite{Bitauld2010}. In previous work \cite{Renema2012,Bitauld2010},
it was shown that such a detector has multiphoton regimes based on
the bias current. The physical mechanism behind these multiphoton
regimes is that at relatively low bias currents, multiple photons
are required to supply a sufficient perturbation for the superconductivity
to be broken. 

The device was cooled in a two-stage pulse-tube / Joule-Thompson cryocooler
to a temperature of approximately 1.2K. The nanodetector was illuminated
using a lensed fiber mounted on cryogenic nanomanipulators. At this
temperature, the overall system detection efficiency for single photons
was $1.5\times10^{-4}$ around our working point at $I_{b}=20\:\mu$A
($I_{c}=29\:\mu$A) . This low efficiency is attributable to the mismatch
between the device active area and the size of the illumination spot.
The device was operated in a voltage bias regime, using a low-noise
voltage source (Yokogawa GS200) in series with a 10 $\Omega$ resistor.
The detector was biased through the DC port of a bias tee, and the
RF signal was amplified in a 45 dB amplifier chain. 

The device was illuminated with a Fianium Supercontinuum laser, whose
pulse duration was specified to be 7 ps. It is crucial for this experiment
that the pulse duration is shorter than the lifetime of an excitation,
which was measured to be several tens of picoseconds \cite{Ilin2000,Rall2010,Zhou}.
If the pulse duration is longer than that, it is possible to have
a pulse which produces two excitations which are far enough apart
in time that one has died out before the second is created; this will
therefore not result in a multiphoton excitation. 

We confirmed that our laser produces coherent states, measuring $g^{(2)}(\tau=0)=0.98\pm0.01$
in a separate experiment. Furthermore, we measured that the intensity
fluctuations in the laser are below 2\%. Hence, the laser is suitable
for tomography \cite{Feito2009}. The detector was illuminated with
narrowband light at wavelengths of 1000 nm, 1300 nm and 1500 nm ($\Delta\lambda$=
10 nm). In our experiment, we vary the intensity and wavelength of
the input light, at various bias currents. At each of these settings,
we record the count rates in a 0.1 s time window and repeat the experiment
10 times per measurement setting. In the current regime investigated
in the present experiment, the detector has negligible dark counts
(< 1 / s). 
\begin{figure}
\includegraphics[width=4cm]{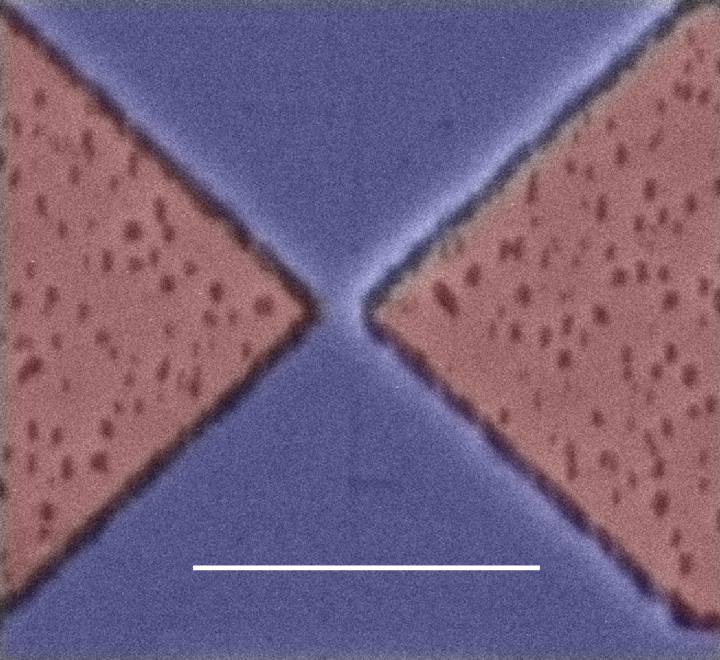}

\caption{False-color SEM image of the detector. The active part of the detector
is the narrow bridge in the centre of the image. The blue parts represent
the thin layer of NbN, the red parts are the GaAs substrate. The scale
bar has a length of 1 $\mu$m.}
\end{figure}

\section{Tomography of multiphoton excitations}

\begin{figure}
\includegraphics{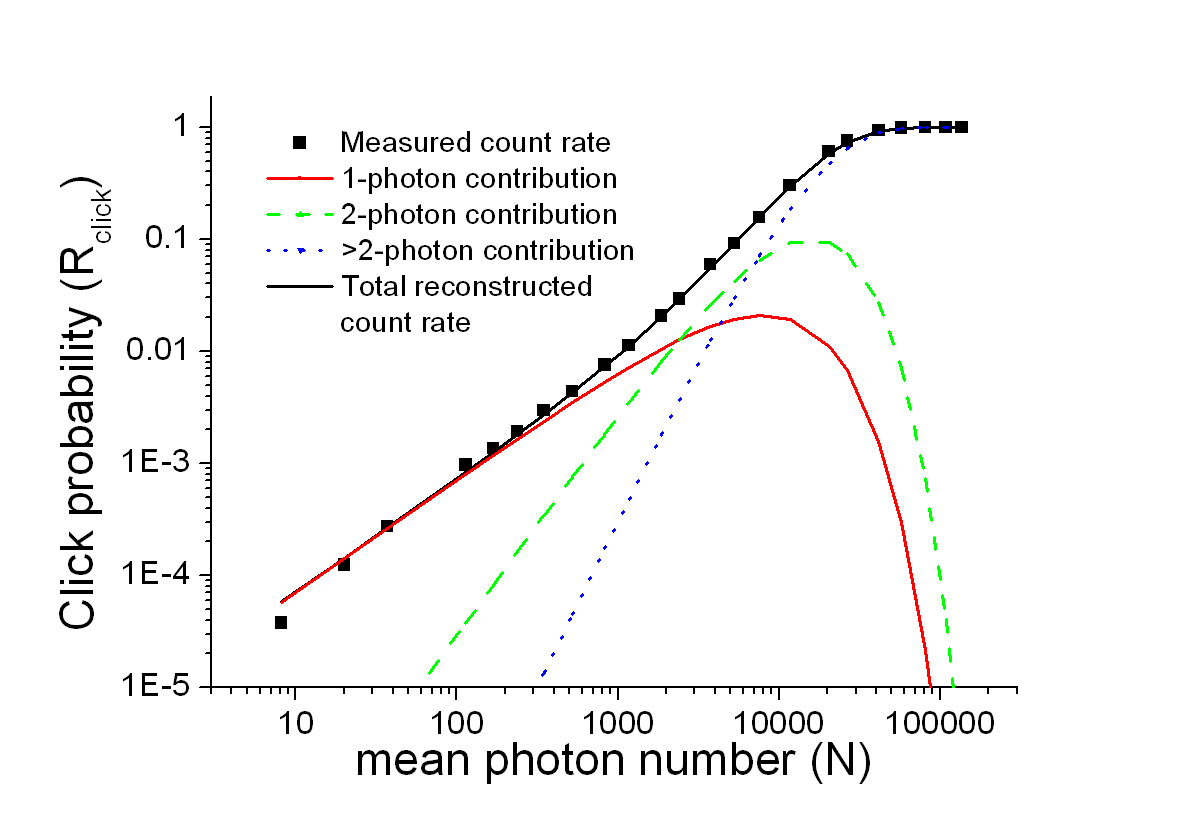}

\caption{Illustration of the tomographic protocol. The black squares indicate
the measured count rate as a function of input power, at $\lambda$
= 1500 nm and $I_{b}=17\:\mu$A. The red (solid) and green (dashed)
lines show the contribution to the count rate of single photons and
photon pairs, respectively. The blue (dotted) line shows the contribution
of higher numbers of photons. The black line shows the sum of all
the photon contributions, indicating that our tomographic reconstruction
succesfully reproduces the observed count rates. From this fit, we
reconstruct the set of detection probabilities $p_{n}$ and the linear
efficiency$\eta$, which together fully describe the behaviour of
the detector.}
\end{figure}

In order to distinguish the effects of the various photon numbers
in the laser pulses, we make use of a sparsity-based tomographic protocol.
We give here a brief summary of this protocol (for a full description,
see ref \cite{Renema2012}, where we introduced this technique). We
illuminate the detector with a range of coherent states, and record
the detection probability $R_{click}$. We make use of two properties
of coherent states: first, that a coherent state under attenuation
remains coherent, second that the decomposition of the coherent state
in the Fock basis is completely determined by the mean photon number,
which can be determined by measuring the intensity %
\footnote{Since we have a phase-insenstive detector, the phase of the coherent
state amplitude is irrelevant, and we set it to zero throughout this
paper for simplicity%
}. 

Each illumination intensity probes the detector with a different linear
combination of photon number states, introducing different combinations
of multiphoton excitations. In particular, we model the detection
probability $R_{click}$ by:
\begin{equation}
R_{click}=1-e^{-\eta N}\sum_{n=0}^{\infty}(1-p_{n})\frac{(\eta N)^{n}}{n!},
\end{equation}
where $\eta$ is the incoupling efficiency and N is the mean photon
number of the incident coherent state. The linear efficiency appears
separately, since our protocol enables us to distinguish linear processes
- such as incoupling to the NbN film - from nonlinear processes\cite{Renema2012}.
The $p_{n}$ are the quantities of interest in further analysis: they
describe the probability of a detection event, given the absorption
of $n$ photons in the active area of the detector. 

Fig. 2 illustrates this protocol as applied to a single experimental
run for a given bias current. We vary the incident power, observe
the detection probability, and apply the tomographic protocol to find
the contributions from the various multiphoton excitations. The black
squares indicate the measured count probability, approaching 1 as
the detector saturates. The red, green and blue lines indicate the
contribution from one photon, two photons and higher photon numbers,
respectively. Only a limited number of multiphoton excitations is
resolvable, and this number depends on bias current. The rest is lumped
into a remainder term containing the limit of high photon numbers
and is not used in further analysis. The fact that at various powers
different multiphoton processes are dominant enables us to recover
them all from a single experiment. Furthermore, since the linear efficiency
$\eta$ only rescales the effective incident photon number, but does
not alter its shape (corresponding to a simple shift in Fig 2), we
are also able to distinguish finite incoupling effects from effects
due to multiphoton excitations.

\section{Results}

Figure 3 shows the reconstructed detection probabilities $p_{n}$,
as a function of bias current and three different wavelengths. For
each wavelength and current, we independently perform the tomographic
procedure outlined in Section III and obtain a full set of parameters
$p_{n}.$We observe that as the current is lowered, the detector makes
a transition from being a one-photon threshold detector to a two-photon
threshold detector, and so on. Furthermore, we observe that the response
curves at different photon numbers and wavelengths have the same shape.
We note that as the excitation energy becomes higher and the photon
number larger, the points on our curves become more scattered, indicating
that the tomography procedure becomes less accurate.

\begin{figure}
\includegraphics[width=8cm]{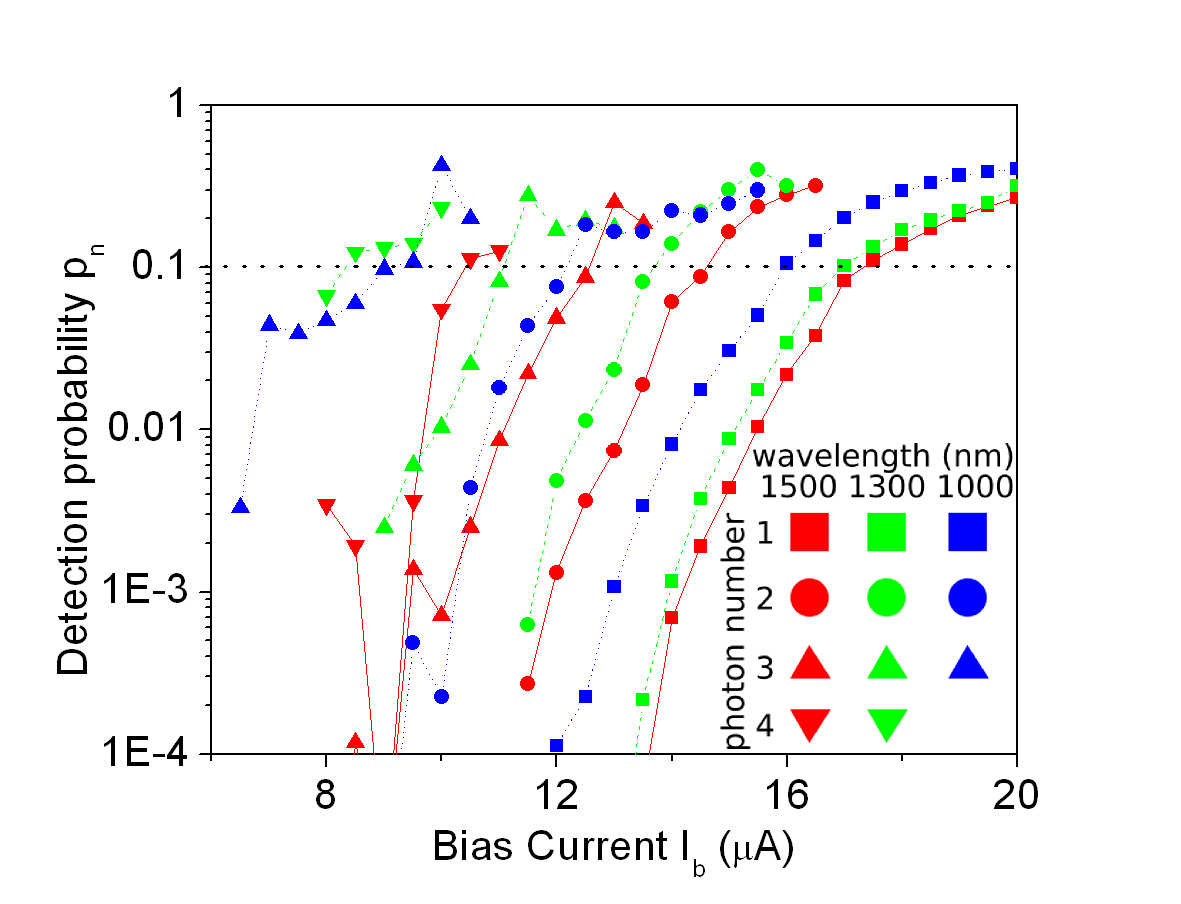}

\caption{Current dependence of the nonlinear parameters $p_{n}$ , as a function
of wavelength and photon number. The probability $p_{n}$ of a detection
event at a given wavelength and photon number is plotted as a function
of the current. The plots are color-coded by wavelength. The shape
of the symbols indicates the photon number (see legend). The connecting
lines are a guide to the eye. The dotted line indicates the threshold
level (p = 0.1) used to obtain Fig. 4. }
\end{figure}

\begin{figure}
\includegraphics{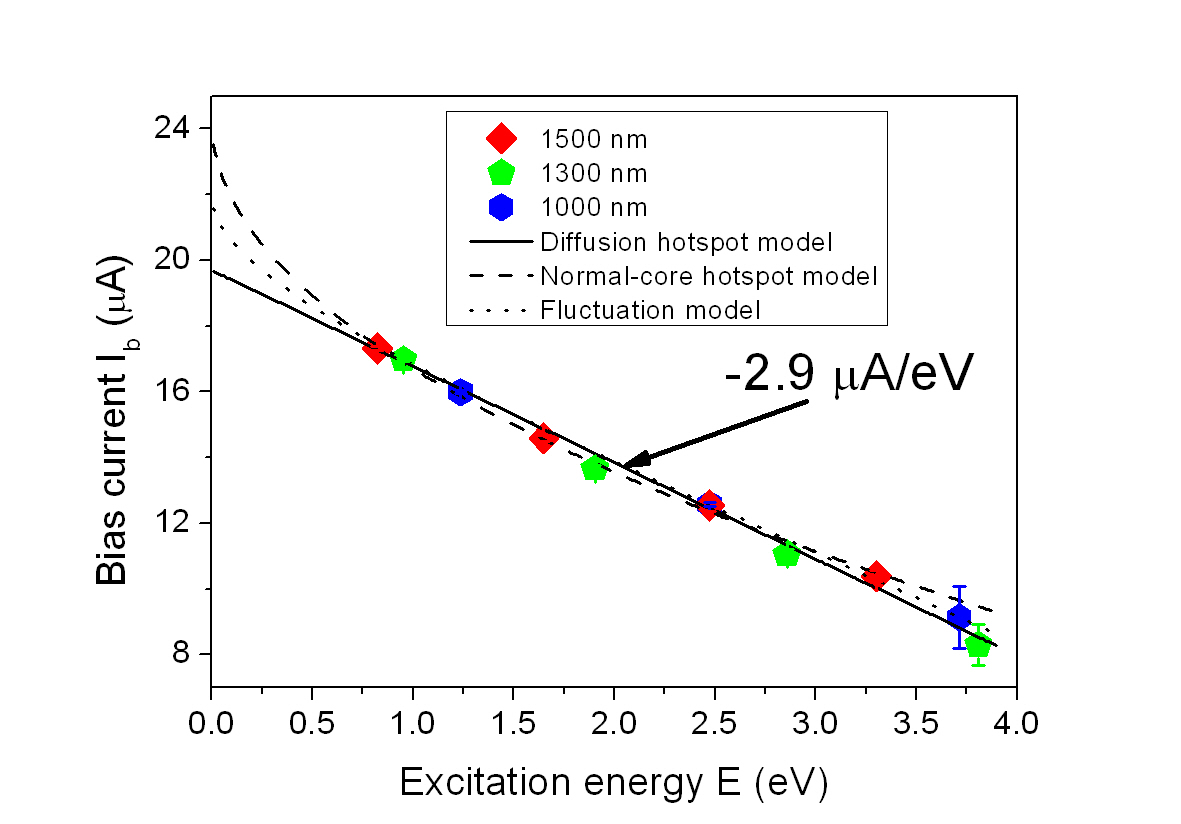}

\caption{Scaling law for the nanodectector. From the data in Fig 3, we find
all points that have $p_{n}(E,I_{b})=0.1$ (indicated by the dotted
line in that figure), where E is the overall excitation energy. In
this graph, we plot the values of $I_{b}$ and $E$ that satisfy this
condition. This graph shows that bias current and overall excitation
energy have an approximately linear dependence. The fact that points
at various photon numbers all fall on the same line demonstrates that
the nanodetector is only sensitive to the overall energy of the excitation.
The three lines show the fits of the three microscopic models to the
data. Apart from the two rightmost points, the errors on these data
points are $\sim$ 100 nA.}
\end{figure}

Figure 4 shows the bias current required to reach a detection probability
of 10\%, as a function of total excitation energy. In order to obtain
this figure, we took a surface of constant $p_{n}(E,I_{b})=0.1$ in
Fig. 3 (indicated by a dotted line), and plotted the bias current
at which the detector has 10\% probability of responding to an energy
$E$, where $E=nh\nu$ is the total energy of the $n$ photons absorbed
by the detector. This figure demonstrates that there is a scaling
law between bias current and overall excitation energy. We determine
the scaling constant$\gamma$ to be $\gamma$ = -2.9 $\pm\:0.1\:\mu\mathrm{A}/\mathrm{eV}\,(=-1.8\times10^{13}\mathrm{\: Wb^{-1}}$in
SI units) for our detector. Furthermore, this figure shows that the
detection probability is independent of the way in which the excitation
is composed of different photons: only the overall energy determines
the detection probability. We note that we have used only a small
fraction of the data present in figure 3 to obtain the data presented
in figure 4. 

We compare three models from literature to our data. We find that
over the range of the experiment, all three models are consistent
with our data. The three models are a hotspot-based model, a hotspot-based
model in which diffusion plays a large role and a fluctiation-assisted
model. For a full description of the three models, see Appendix. These
models distinguish themselves not only by different detection mechanisms,
but also by different scaling between bias current and energy at constant
detection efficiency. 

\begin{figure}
\includegraphics{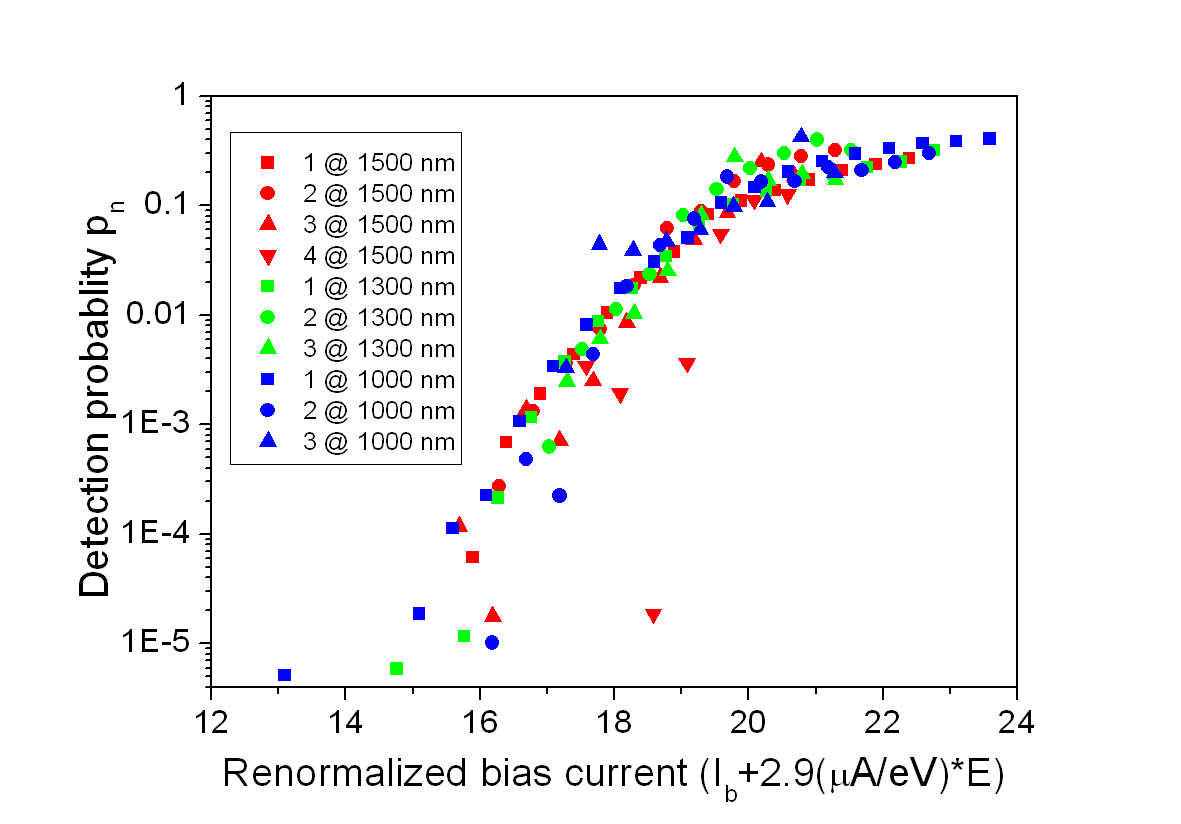}

\caption{Universal response curve for the nanodetector. To obtain these curves,
we rescale the curves reported in Fig. 3 by the scaling law demonstrated
in Fig. 4. }
\end{figure}

Figure 5 presents the main result of our paper: a universal detection
curve for a single line-segment of an NbN SSPD. In Fig. 5, we apply
the scaling law, which was derived from the points around $p=0.1$
to the entire data set. We find that all the curves of detection probability
as a function of rescaled bias current superimpose over more than
3 orders of magnitude in the detection probability. This shows that
the photoresponse of our detector depends only on this specific combination
$I_{b}+\gamma E$ of bias current and excitation energy. We stress
that this universal curve can only be obtained through detector tomography,
which allows separation of the effects of multiphoton excitation and
finite linear efficiency.

The data presented in Fig. 5 shows that the scaling behaviour which
we obtained at $p_{i}=0.1$ in figure 4 is universal for all values
of $p$. Since we have used only the points in figure 3 which lie
around to obtain the result in figure 4, we do not a priori expect
the curves to superimpose when we apply the scaling law to the entire
dataset. In such a procedure, only the points which are used to obtain
the scaling factor are guaranteed to superimpose. Since the curve
is universal over more than 4 orders of magnitude in the detection
probability, we have demonstrated that our results are independent
of the arbitrary choice of the 10\% criterion. The criterion only
matters for the accuracy with which the curves can be superimposed:
we find from theoretical simulations that the tomographic reconstruction
is most accurate between $p_{n}=0.1$ and $p_{n}=10^{-4}.$ This justifies
the choice of our criterion.

\section{Discussion\ }

In this section, we first compare our experimental method with that
of previous studies on SSPDs. Then, we discuss our experimental findings
on the universal curve, quasiparticle conversion efficiency. Lastly,
we discuss the phenomenon of dark counts in our detector.

\subsection{\emph{Comparison with previous work} }

Previous investigations of the SSPDs detection mechanism were of a
semiclassical nature, where only the efficiency and dark count rate
were measured. By observing the exponent of the power-law dependence
of count rate on input power, one can also infer the photon number
detection regime semiclassically. However, such a characterization
is limited to the observation that the detector is operating in a
particular detection regime; a measurement of the $p_{i}$ (i.e. how
strongly the detector is in a particular regime) requires detector
tomography. Since the width of each multiphoton regime is $\sim$
2 $\mu$A for our experiment, the accuracy of the semiclassical method
is rather limited. In order to characterize multiphoton processes
beyond that resolution, detector tomography is an absolute requirement.

Most previous work focussed on meander detectors, which is the geometry
that is normally used in practical applications of SSPDs. In a meander,
two photons that are absorbed in different places along the wire do
not constitute a two-photon event, yet they may still produce one-photon
events individually. By using a nanodetector, we sidestep any question
of how the photons distribute themselves along the length of the wire,
which was a major issue in measuring multiphoton effects in meander-type
SSPDs \cite{Akhlaghi2009,Akhlaghi2009a}. 

Our present work probes the detection mechanism at various energies
simultaneously. We are insensitive to incoupling losses, since they
affect the various multiphoton processes equally. Furthermore, since
we can perform excitations at different energies with the same wavelength,
we are insensitive to any wavelength-dependent effects in the experiment,
including wavelength-dependent absorption in the NbN layer.

\subsection{Universal curve}

The universal curve which we demonstrate in Fig. 5 is not predicted
by any of current SSPD photodetection models. Typically, such models
focus on calculating a single threshold bias current $I_{th}$, above
which the energy of a photon is large enough to deterministically
break the superconductivity. Above that current, the efficiency of
the detector should be constant. We have shown in the present work
that scaling behaviour extends not just to a single threshold current,
but to all combinations of currents and excitations in the present
experiment. Scaling behaviour applies whether one is in the regime
of high effiency or not. This points to the fact that a single theory
should describe detection events in SSPDs, both in the high and low-efficiency
regimes.

\subsection{Quasiparticle conversion efficiency}

The fact that only the overall energy of the excitation determines
the response of the photodetector can be interpreted in terms of the
cascade process that is generated by the initial excitation. This
process, which is thought to involve both electrons and phonons in
the film, and in which the mutual exchange of energy between the electron
and phonon subsystem plays a key role, is still poorly understood.
In the present work, we probe this cascade process with different
initial excitations, and show that it is only the overall energy which
determines the total number of quasiparticles which are produced at
the superconducting band-edge. The fact that four excitations of a
quarter of the energy produce the same number of QP as a single excitation
with the full energy is evidence of the fact that the conversion efficiency
by which the energy of the first QP is distributed over many others
is independent of the initial energy.

\subsection{Dark counts}

We now turn to the phenomenon of dark counts. The most straightforward
model is the following: one simply considers a dark count as an excitation
with E = 0. Extrapolating the linear scaling law from Fig. 4 to $E=0$
yields a current of 19 $\mu$A. However, at this current we do not
observe a dark count probability of 10\% as one would expect from
the simple model; we only observe appreciable dark counts around the
critical current of 29 $\mu$A. The same discrepancy applies to the
other two models. We can therefore say that the picture of a dark
count as a zero-energy photodetection event is not supported by our
data for any current detection model of SSPDs. The anomalous behaviour
of dark counts is a reminder of the danger of assuming a detection
model, further demonstrating the relevance of our tomographic method.
In particular in this case, the tomographic method gives the first
hints of substantial differences in detection mechanism between dark
counts and light counts. We note that the nature of dark counts is
still open to debate \cite{Bulaevskii2011,Gurevich2012,Gurevich2008}.

\subsection{Outlook}

The present work opens up the possibility of testing the various models
of photodetection. This could be done by performing the present experiment
in the mid-infrared. For this energy range the predictions of the
various models differ significantly (see Fig. 4). For example, at
an excitation wavelength of 5 $\mu$m, corresponding to 0.25 eV, the
difference between the predictions of the various models is easily
measurable; it is of the order of 1 $\mu$A.

Multiphoton excitation has the practical advantage that the bandwidth
of energy excitations which is offered can be extended by a factor
equal to the number of photons in the highest excitation (in our case,
4). This has applications in the situation where light of a particular
wavelength is difficult to couple onto a cryogenic sample. In particular,
the present work opens up the possibility of studying NbN detector
behaviour in an energy range that corresponds to the near and medium
UV range, using visible and NIR optics.

In a previous publication \cite{Zhou}, we have introduced the notion
of the nonlinear response function (NRF) $\eta(I_{b},C),$ which measures
the instantaneous detection probability, given that a bias current
of $I_{b}$ is present, and that there are $C$ quasiparticles in
the detector. The overall detection probability is then given by R
=$\int_{t}\eta(t)I(t)dt$, where $I$ is the instantaneous intensity.
This function can be probed by various means such as a pump-probe
experiment. The description in terms of a NRF is well-matched to a
tomographic experiment, as both are model-independent descriptions. 

The holy grail of tomographic research on SSPDs would be to find the
instantaneous detection probability as a function of the number of
quasiparticles present at that instant. In the present experiment,
we have achieved a step towards this goal: we have demonstrated the
NRF to be of the form $\eta(I_{b}+\gamma E)$ over the energy range
of the experiment, for short-pulse excitations.

\section{Conclusions}

In conclusion, we have studied the physics of photodetection in a
superconducting single photon detector. We have shown that the detection
is based on the overall energy of the excitation. Furthermore, we
have demonstrated a scaling law between overall excitation energy
and bias current. From this, we find a universal response curve that
depends only on a given combination of bias current and excitation
energy. Thereby, we have shown that the known behaviour of the detector
extends into the multiphoton range. These results demonstrate that
the tomographic method is a useful tool for investigating the fundamenal
physics of detection events in NbN SSPDs. 
\begin{acknowledgments}
We thank D. Sahin for providing the SEM image of the detector. We
thank Prof. G. Goltsman, Prof. P. Kes, Prof. J Aarts, Prof. R. Gill,
Q. Wang and R. J. Rengelink for useful discussions. This work is part
of the research programme 'Nanoscale Quantum Optics' of the Foundation
for Fundamental Research on Matter (FOM), which is financially supported
by the Netherlands Organisation for Scientific Research (NWO) and
is also supported by NanoNextNL, a micro- and nanotechnology program
of the Dutch Ministry of Economic Affairs, Agriculture and Innovation
(EL\&I) and 130 partners, and by the European Commission through FP7
project Q-ESSENCE (contract No. 248095).\appendix
\end{acknowledgments}

\section*{Appendix: detection models in SSPDs}

\subsection{Introduction}

While big strides have been taken \cite{Engel,Gurevich2012,Gurevich2008,Bulaevskii2011}
in understanding the fundamental physics of these detectors, many
details of the detection mechanism in such detectors are still unknown.
After the absorption of a photon, it is thought that the resulting
high-energy electron destroys Cooper pairs that carry the bias current,
producing a cloud of quasi-particles \cite{Semenov2005,Hofherr2010}.
This process results in a breakdown of the superconductivity, resulting
in a normal cross-section. After such a resistive barrier has formed,
the kinetic inductance of the device drives Joule heating in the normal
state area \cite{Kerman2006}, causing it to grow. After that, the
current drops to a negligible level \cite{Kerman2007} and the interplay
between the cooling of the device and the restoration of the current
determines whether the device returns to its previous state, ready
to detect another photon. 

Currently, there is no consensus on a microscopic model for the detection
event in superconducting single photon detectors. Below, we briefly
introduce three microscopic models for detection events in the SSPD
from literature. We refer to these three as the normal-core hotspot
model\emph{, }the\emph{ }diffusion hotspot model\emph{ }and\emph{
}the\emph{ }fluctuation model. Each of these models predicts that
the energy and current are exchangeable through a given scaling law,
and their prediction will be compared with the experimental data below.

\subsection{Three models}

The \emph{normal-core hotspot model} was introduced in the original
papers reporting photodetection with SSPDs \cite{Goltsman2001,Semenov2001}.
In this model, it is assumed that the photon absorption creates a
normal core inside the material. Current is then diverted around this
core. If the current locally exceeds the critical current, superconductivity
is destroyed and a normal state slab is created, resulting in a detection
event. In this model, the current required to pinch off the entire
channel for a given input energy scales as the square root of the
energy, since the hotspot is assumed to be a cylindrical object inside
the wire. For a given energy, the bias current $I_{b}$ needed to
achieve maximal detection efficiency is then given by: 
\begin{equation}
E=\frac{w^{2}}{C^{2}}(1-I_{b}/I_{c})^{2},
\end{equation}
where E is the energy of the photon, $I_{b}$ is the bias current
and $I_{c}$ is the critical current. $w$ is the width of the wire
and $C$ is a scaling constant, which is defined in this way for consistency
with previous work \cite{Suzuki2011}. 

In the \emph{diffusion-hotspot model}, which was introduced later
as a refinement of the original normal-core hotspot model, the role
of the diffusion of quasiparticles is taken into account, as well
as the reduction of the critical current due to the quasiparticles
\cite{Engel,Semenov2005}. In this model, the relevant lengthscale
is given by the diffusion length over a time characteristic for the
cascade of quasiparticles. This expression, which was first derived
in Ref. \cite{Semenov2005}, predicts linear scaling between bias
current and cutoff energy:

\begin{equation}
E=E_{0}(1-I_{b}/I_{c}),
\end{equation}
where $E_{0}$ is an energy scale \cite{Hofherr2010}. 

\emph{Fluctuation model }The previous two models predict a sharp cutoff
of the detection efficiency as a function of photon energy, which
is not observed in experiments\cite{Hofherr2010}. In order to explain
the observed detection probability beyond the cutoff energy, fluctuation-assisted
detection models have been proposed \cite{Bulaevskii2011,Semenov2005,Gurevich2008,Gurevich2012}.
All of these have in common that the role of the photon in the detection
process is to depress the superconducting gap. Subsequently, a thermally
activated fluctuation occurs, such as the depairing of a vortex-antivortex
pair (VAP) \cite{Semenov2008} in the superconductor or the crossing
of a single vortex. This fluctuation must overcome an energy barrier
$E(\Delta,I_{b})$ \cite{Bulaevskii2011}. Expressions for such energy
barriers typically contain a gap-dependent energy scale and a current-dependent
geometric factor \cite{Hofherr2010}. The specifics of the geometric
factor depends on the precise fluctuation process. Both for the VAP
model and for the single-vortex crossing model, we can linearize the
current-dependence of the geometric factor over the range of currents
that was used in the experiment, to obtain: 

\begin{equation}
A=(\Delta-\alpha\sqrt{E})(I_{0}-\beta I_{b}),
\end{equation}
where the constants $I_{0}$ and $\beta$ are known from the linearization
of the geometric factor, and A and $\alpha$ are the experimental
fit parameters. Such a model predicts a hyperbolic interchange between
energy and current.

We note the following subtle point: to obtain figure 4, one must introduce
an arbitrary cutoff criterion, which we have chosen to be $p_{i}=0.1$
(see main text). This introduces an offset in the parameters of each
model, in particular in the parameter which describes the behaviour
at E = 0. We have constructed the expressions for each model such
that the choice of cutoff manifests itself as an offset in a single
parameter, namely the current scale appropriate for that model. The
parameters reported below are independent of the choice of the cutoff
criterion.

\subsection{Comparison with experimental results}

We apply the three detection models to the results in Figure 4, and
compare the results with the values from literature. For the normal-core
hotspot model, we find $C=47\:\pm1$ eV$^{-1/2}$/ nm, which should
be compared to the values of $C=11-20$ eV$^{-1/2}$/ nm found in
other experiments \cite{Verevkin2002}. For the diffusion hotspot
model, we apply the expression from Ref. \cite{Hofherr2010}, to find
a theoretical value of $\gamma$ = -2.5 $\mu$A/eV for our sample
and $\gamma$ = -3.5 $\mu$A/eV for the samples in that reference,
which should be compared with the value of $\gamma$ = -2.9 $\pm$
0.1 $\mu$A/eV obtained experimentally. For the fluctuation model,
we find $\alpha=2.8\times10^{-4}\pm0.05\times10^{-4}\sqrt{\mathrm{eV}},$
which should be compared to a literature value of $\alpha=6\times10^{-4}\sqrt{\mathrm{eV}},$
for the experiment reported in Ref. \cite{Semenov2008}. We note,
however, that comparisons between different detectors are problematic.
In particular, the conversion efficiency of the initial excitation
to quasiparticles at the gap edge is a free parameter which varies
from detector to detector\cite{Hofherr2010}. 

The error analysis on the quantities given in the previous paragraph
was based on the 50 nA accuracy of the current readout of our experiment,
combined with error propagation on the interpolation formula used
to obtain the intersection with the line $p_{i}=0.1$. For low $i$,
the former error dominates. At higher $i$, we are limited by the
quality of our tomographic reconstruction. We calculate $\chi^{(2)}$
per degree of freedom to be 2.2, 2.9, and 2.1 for the normal-core
hotspot, diffusion hotspot and fluctiation models, respectively. These
numbers do not enable us to conclusively rule out any of the models.

In the normal-core hotspot model, multiphoton events can only occur
if the various normal cores are created at the cross-section of the
wire. Given the relative size of our detector and the size of the
normal cores in this hotspot model, we would expect to see reduced
count rates at high photon numbers due to the low probability of several
photons being absorbed at exactly the same point in our detector.
The fact that we do not observe such reduced count rates suggests
that in our experiment, the size of the hotspot is comparable to the
size of our detector. 
\begin{acknowledgments}
\bibliographystyle{plain}
\bibliography{paper4}

\begin{thebibliography}{10}

\bibitem{Akhlaghi2009}
Mohsen~K Akhlaghi, A~Hamed Majedi, and Jeff~S Lundeen.
\newblock {Nonlinearity in Single Photon Detection : Modeling and Quantum
  Tomography}.
\newblock {\em Opt. Express}, 19:21305--21312, 2011.

\bibitem{Akhlaghi2009a}
Mohsen~Keshavarz Akhlaghi and Amir~Hamed Majedi.
\newblock {Semiempirical Modeling of Dark Count Rate and Quantum Efficiency of
  Superconducting Nanowire Single-Photon Detectors}.
\newblock {\em IEEE Trans. Appl. Supercond.}, 19(3):361--366, 2009.

\bibitem{Annunziata2009}
A.J. Annunziata, D.F. Santavicca, J.D. Chudow, L.~Frunzio, M.J. Rooks,
  A.~Frydman, and D.E. Prober.
\newblock {Niobium Superconducting Nanowire <newline/>Single-Photon Detectors}.
\newblock {\em IEEE Transactions on Applied Superconductivity}, 19(3):327--331,
  June 2009.

\bibitem{Bitauld2010}
David Bitauld, Francesco Marsili, Alessandro Gaggero, Francesco Mattioli,
  Roberto Leoni, Saeedeh Jahanmirinejad, Francis L\'{e}vy, and Andrea Fiore.
\newblock {Nanoscale optical detector with single-photon and multiphoton
  sensitivity.}
\newblock {\em Nano Lett.}, 10(8):2977--81, August 2010.

\bibitem{Boroson2009}
D.~M. Boroson, J.~J. Scozzafava, D.~V. Murphy, and B.~S. Robinson.
\newblock {The Lunar Laser Communications Demonstration (LLCD)}.
\newblock {\em 2009 Third IEEE International Conference on Space Mission
  Challenges for Information Technology}, (Llcd):23--28, July 2009.

\bibitem{Bulaevskii2011}
L.~Bulaevskii, M.~Graf, C.~Batista, and V.~Kogan.
\newblock {Vortex-induced dissipation in narrow current-biased thin-film
  superconducting strips}.
\newblock {\em Phys. Rev. B}, 83(14):144526, April 2011.

\bibitem{Collins2007}
R.J. Collins, R.H. Hadfield, V.~Fernandez, S.W. Nam, and G.S. Buller.
\newblock {Low timing jitter detector for gigahertz quantum key distribution}.
\newblock {\em Electron. Lett.}, 43(3):180, 2007.

\bibitem{Dorenbos2008}
S.~N. Dorenbos, E.~M. Reiger, U.~Perinetti, V.~Zwiller, T.~Zijlstra, and T.~M.
  Klapwijk.
\newblock {Low noise superconducting single photon detectors on silicon}.
\newblock {\em Appl. Phys. Lett.}, 93(13):131101, 2008.

\bibitem{Engel2012}
A.~Engel, A.~Aeschbacher, K.~Inderbitzin, A.~Schilling, K.~Il\"{'}in,
  M.~Hofherr, M.~Siegel, A.~Semenov, and H.-W. H\"{u}bers.
\newblock {Tantalum nitride superconducting single-photon detectors with low
  cut-off energy}.
\newblock {\em Appl. Phys. Lett.}, 100(6):062601, 2012.

\bibitem{Engel}
A.~Engel, Kevin Inderbitzin, Andreas Schilling, Robert Lusche, Alexei Semenov,
  Dagmar Henrich, Matthias Hofherr, Konstantin Il, and Michael Siegel.
\newblock {Temperature-dependence of detection efficiency in NbN and TaN
  SNSPD}.
\newblock 2012.

\bibitem{Feito2009}
A.~Feito, J~S Lundeen, H~Coldenstrodt-Ronge, J~Eisert, M~B Plenio, and I~A
  Walmsley.
\newblock {Measuring measurement: theory and practice}.
\newblock {\em New J. Phys.}, 11(9):093038, September 2009.

\bibitem{Gaggero2010}
A.~Gaggero, S.~Jahanmirinejad, F.~Marsili, F.~Mattioli, R.~Leoni, D.~Bitauld,
  D.~Sahin, G.~J. Hamhuis, R.~N\"{o}tzel, R.~Sanjines, and A.~Fiore.
\newblock {Nanowire superconducting single-photon detectors on GaAs for
  integrated quantum photonic applications}.
\newblock {\em Appl. Phys. Lett.}, 97(15):151108, 2010.

\bibitem{Goltsman2001}
G.~N. Goltsman, O.~Okunev, G.~Chulkova, A.~Lipatov, A.~Semenov, K.~Smirnov,
  B.~Voronov, A.~Dzardanov, C.~Williams, and Roman Sobolewski.
\newblock {Picosecond superconducting single-photon optical detector}.
\newblock {\em Appl. Phys. Lett.}, 79(6):705, 2001.

\bibitem{Gurevich2008}
A.~Gurevich and V.~Vinokur.
\newblock {Size Effects in the Nonlinear Resistance and Flux Creep in a Virtual
  Berezinskii-Kosterlitz-Thouless State of Superconducting Films}.
\newblock {\em Phys. Rev. Lett.}, 100(22):227007, June 2008.

\bibitem{Gurevich2012}
A.~Gurevich and V.~Vinokur.
\newblock {Comment on Vortex-assisted photon counts and their magnetic field
  dependence in single-photon superconducting detectors}.
\newblock {\em Phys. Rev. B}, 86(2):026501, July 2012.

\bibitem{Hadfield2006}
Robert~H. Hadfield, Jonathan~L. Habif, John Schlafer, Robert~E. Schwall, and
  Sae~Woo Nam.
\newblock {Quantum key distribution at 1550nm with twin superconducting
  single-photon detectors}.
\newblock {\em Appl. Phys. Lett.}, 89(24):241129, 2006.

\bibitem{GisinnatureP}
M.~Halder, A.~Beveratos, M.~Gisin, V.~Scarani, C.~Simon, and H.~Zbinden.
\newblock {\em Nat. Phys.}, 3:692--695, 2007.

\bibitem{Hofherr2010}
M.~Hofherr, D.~Rall, K.~Ilin, M.~Siegel, A.~Semenov, H.-W. H\"{u}bers, and
  N.~A. Gippius.
\newblock {Intrinsic detection efficiency of superconducting nanowire
  single-photon detectors with different thicknesses}.
\newblock {\em J. Appl. Phys.}, 108(1):014507, 2010.

\bibitem{Ilin2000}
K.S. Ilin, M.~Lindgren, M.~Currie, A.D. Semenov, G.N Goltsman, R.~Sobolevski,
  S.I. Cherednichenko, and E.M. Gershenzon.
\newblock {\em Appl. Phys. Lett.}, 76:2752, 2000.

\bibitem{Kerman2006}
Andrew~J. Kerman, Eric~A. Dauler, William~E. Keicher, Joel K.~W. Yang, Karl~K.
  Berggren, G.~Goltsman, and B.~Voronov.
\newblock {Kinetic-inductance-limited reset time of superconducting nanowire
  photon counters}.
\newblock {\em Appl. Phys. Lett.}, 88(11):111116, 2006.

\bibitem{Kerman2007}
Andrew~J. Kerman, Eric.~A. Dauler, Joel K.~W. Yang, Kristine~M. Rosfjord, Vikas
  Anant, Karl~K. Berggren, Gregory~N. Goltsman, and Boris~M. Voronov.
\newblock {Constriction-limited detection efficiency of superconducting
  nanowire single-photon detectors}.
\newblock {\em Appl. Phys. Lett.}, 90(10):101110, 2007.

\bibitem{Marsili}
F.~Marsili, V.~B. Verma, J.~A. Stern, S.~Harrington, A.~E. Lita, T.~Gerrits,
  I.~Vayshenker, and B.~Baek.
\newblock {Detecting Single Infrared Photons with 93 System Efficiency}.
\newblock 2012.

\bibitem{Mattioli2012}
Francesco Mattioli, Mikkel Ejrnaes, Alessandro Gaggero, Alesandro Casaburi,
  Roberto Cristiano, Sergio Pagano, and Roberto Leoni.
\newblock {Large area single photon detectors based on parallel configuration
  NbN nanowires}.
\newblock {\em J. Vac. Sci. Technol.}, 30(031204), 2012.

\bibitem{Mohan2008}
Nishant Mohan, Olga Minaeva, Gregory~N Goltsman, Magued~B Nasr, Bahaa~E Saleh,
  Alexander~V Sergienko, and Malvin~C Teich.
\newblock {Photon-counting optical coherence-domain reflectometry using
  superconducting single-photon detectors.}
\newblock {\em Opt. Express}, 16(22):18118--30, October 2008.

\bibitem{Rall2010}
D~Rall, P~Probst, M~Hofherr, S~W\"{u}nsch, K~Il'in, U~Lemmer, and M~Siegel.
\newblock {Energy relaxation time in NbN and YBCO thin films under optical
  irradiation}.
\newblock {\em Journal of Physics: Conference Series}, 234(4):042029, June
  2010.

\bibitem{Renema2012}
J~J Renema, G~Frucci, Z~Zhou, F~Mattioli, A~Gaggero, R~Leoni, M~J A~De Dood,
  A~Fiore, and M~P~Van Exter.
\newblock {Modified detector tomography technique applied to a superconducting
  multiphoton nanodetector}.
\newblock {\em Opt. Express}, 20(3):2806--2813, 2012.

\bibitem{RenemaPRA}
J.J. Renema, G.~Frucci, M.J.A. de~Dood, R~Gill, A~Fiore, and M.P. van Exter.
\newblock {\em Phys. Rev. A}, 86:062113, 2012.

\bibitem{Semenov2005}
A.~Semenov, A.~Engel, H.-W. H\"{u}bers, K.~Il'in, and M.~Siegel.
\newblock {Spectral cut-off in the efficiency of the resistive state formation
  caused by absorption of a single-photon in current-carrying superconducting
  nano-strips}.
\newblock {\em Euro. Phys. J. B}, 47(4):495--501, November 2005.

\bibitem{Semenov2001}
Alex~D Semenov and Alexander~A Korneev.
\newblock {Quantum detection by current carrying superconducting ®lm}.
\newblock {\em Physica C}, 351:349--356, 2001.

\bibitem{Semenov2008}
Alexei~D. Semenov, Philipp Haas, Heinz-Wilhelm H\"{u}bers, Konstantin Ilin,
  Michael Siegel, Alexander Kirste, Thomas Schurig, and Andreas Engel.
\newblock {Vortex-based single-photon response in nanostructured
  superconducting detectors}.
\newblock {\em Physica C}, 468(7-10):627--630, April 2008.

\bibitem{Stevens2010}
Martin~J Stevens, Burm Baek, Eric~A Dauler, Andrew~J Kerman, Richard~J Molnar,
  Scott~A Hamilton, Karl~K Berggren, Richard~P Mirin, and Sae~Woo Nam.
\newblock {High-order temporal coherences of chaotic and laser light.}
\newblock {\em Opt. Express}, 18(2):1430--7, January 2010.

\bibitem{Suzuki2011}
Koji Suzuki, Shigetomo Shiki, Masahiro Ukibe, Masaki Koike, Shigehito Miki,
  Zhen Wang, and Masataka Ohkubo.
\newblock {Hot-Spot Detection Model in Superconducting Nano-Stripline Detector
  for keV Ions}.
\newblock {\em Applied Physics Express}, 4(8):083101, July 2011.

\bibitem{Verevkin2002}
A.~Verevkin, J.~Zhang, Roman Sobolewski, A.~Lipatov, O.~Okunev, G.~Chulkova,
  A.~Korneev, K.~Smirnov, G.~N. Gol'tsman, and A.~Semenov.
\newblock {Detection efficiency of large-active-area NbN single-photon
  superconducting detectors in the ultraviolet to near-infrared range}.
\newblock {\em Appl. Phys. Lett.}, 80(25):4687, 2002.

\bibitem{Zhang2003}
J.~Zhang, W.~Slysz, A.~Pearlman, A.~Verevkin, Roman Sobolewski, O.~Okunev,
  G.~Chulkova, and G.~Goltsman.
\newblock {Time delay of resistive-state formation in superconducting stripes
  excited by single optical photons}.
\newblock {\em Phys. Rev. B}, 67(13):132508, April 2003.

\bibitem{Zhou}
Zili Zhou, Giulia Frucci, Francesco Mattioli, Alessandro Gaggero, Saeedeh
  Jahanmirinejad, Thang~Ba Hoang, and Andrea Fiore.
\newblock {Ultrasensitive N -photon interferometric autocorrelator}.
\newblock 2012.

\bibitem{Zinoni2007}
C.~Zinoni, B.~Alloing, L.~H. Li, F.~Marsili, A.~Fiore, L.~Lunghi, A.~Gerardino,
  Yu.~B. Vakhtomin, K.~V. Smirnov, and G.~N. Goltsman.
\newblock {Single-photon experiments at telecommunication wavelengths using
  nanowire superconducting detectors}.
\newblock {\em Appl. Phys. Lett.}, 91(3):031106, 2007.

\end{thebibliography}
\end{acknowledgments}

\end{document}